\documentclass[twoside,aps,showpacs,floatfix]{revtex4}

\usepackage{amsmath,amssymb,graphicx,dsfont}
\usepackage[latin1]{inputenc}

\newcommand{\al}{\alpha}
\newcommand{\vs}{\vec{S}}
\newcommand{\vsi}{\vec{\sigma}}
\newcommand{\rh}{\hat{r}}
\newcommand{\atanh}{\text{artanh}}

\begin{document}

\title{Replica theory for Levy spin glasses \footnote{Dedicated to
    Professor Thomas Nattermann on the occasion of his
    60$^\mathrm{th}$ anniversary.}}

\author{K. Janzen}
\email{janzen@theorie.physik.uni-oldenburg.de} 
\author{A. K. Hartmann}
\email{a.hartmann@uni-oldenburg.de} 
\author{A. Engel}
\email{engel@theorie.physik.uni-oldenburg.de} 
\affiliation{Institut f\"ur Physik, Carl-von-Ossietzky-Universtit\"at,
     26111 Oldenburg, Germany}


\begin{abstract}
Infinite-range spin-glass models with Levy-distributed interactions
show a spin-glass transition with similarities to both the
Sherrington-Kirkpatrick model and to disordered spin systems on
finite connectivity random graphs. Despite the diverging moments of the
coupling distribution the transition can be analyzed within the
replica approach by working at imaginary temperature. Within the
replica-symmetric approximation a self-consistent equation for the
distribution of local fields is derived and from the instability of the
paramagnetic solution to this equation the glass-transition
temperature is determined. The role of the percolation of rare strong
bonds for the transition is elucidated. The results partly agree and
partly disagree with those obtained within the cavity
approach. Numerical simulations using parallel tempering are in
agreement with the transition temperatures found.
\end{abstract}

\maketitle

\section{Introduction}
Spin glasses have been one of the most prominent models for disordered
systems since the classical paper by Edwards and Anderson
\cite{EdAn}. They are built from simple degrees of freedom interacting
via random couplings \cite{BiYo}. The ensuing interplay between
disorder and frustration gives rise to peculiar static and 
dynamic properties which made spin-glasses paradigms for complex
systems with competing interactions. The concepts and techniques
developed for their theoretical understanding became useful also 
in the quantitative analysis of problems from algorithmic complexity
\cite{RemiNat,MPZ,HaWe}, game theory \cite{DiOp,EnBe}, artificial
neural networks \cite{Amit,EnvB}, and cryptography \cite{KaKi}.  

A comprehensive understanding of spin glasses is so far possible on
the mean-field level only. Different models of mean-field spin glasses
have been introduced and analyzed over the years. The
Sherrington-Kirkpatrick (SK) model \cite{SK} was designed as
generalization of the Weiss model of ferromagnetism. It is the most
popular completely connected spin glass model in which each spin
interacts with all ${\cal O}(N)$ other spins via {\it weak} couplings 
$J_{ij}={\cal O}(N^{-1/2})$. The central limit theorem can then be
invoked to determine the statistical properties of the local fields
and in the simplest situation the distribution of these fields is
Gaussian and can be characterized by a single scalar order parameter. 
The details of the spin glass transition
and the intricate nature of the low-temperature phase of this model
have been thoroughly elucidated within the framework of the celebrated
Parisi solution \cite{Parisi}. Quite recently the main features of
this solution were established in a mathematically rigorous way
\cite{Talagrand}. 

The variety of mean-field models for spin glasses is, however, by far
not exhausted by SK-like systems. The Viana-Bray (VB) model
\cite{ViBr} and, more generally, spin glasses on finite-connectivity
graphs \cite{KaSo,MePa} combine finite coordination number with
mean-field behaviour. Here each spin interacts with a finite
number of randomly selected other spins through {\it strong} bonds 
$J_{ij}={\cal O}(1)$. Accordingly the distribution of local fields is 
not Gaussian and has to be characterized by all its moments. The
analysis of these systems is therefore technically more involved and 
already in the simplest (replica symmetric) description infinitely
many order parameters (or equivalently an order parameter function)
have to be introduced. Models of this type often arise in the
analysis of complex optimization problems with methods from
statistical mechanics \cite{RemiNat}. 

>From the technical perspective two different methods were developed to
analyze spin glasses within the framework of equilibrium statistical
mechanics. The replica method \cite{Edwards,EdAn,MPV} centers around
averages of integer moments of the partition function of the system. 
Its crucial step consists in the analytical continuation of the
results for the $n$-th moment of the partition function from integer
to real $n$ and the final limit $n\to 0$. Complementary, the cavity
method \cite{ParLH,MPV,MePa01} builds on 
the clustering property of equilibrium states and the stability of the
thermodynamic limit $N\to\infty$. Here one considers a system with
$N$ spins, adds one additional spin with its couplings to the system
and derives self-consistency relations that stem from the fact that
the statistical properties of the $N$ and the $N+1$ spin system should
be similar to each other. Both the SK and the VB model have
been analyzed using the replica as well as the cavity method. 

Also from the numerical point of view, spin glasses are a challenging
problem. Due to the frustrated interactions and since no efficient cluster
algorithm exists, it is very
hard to equilibrate samples at temperatures below the transition
temperature. Hence, other schemes like
parallel tempering are used frequently, but still sizes of the order of
$N=1000$ spins are typically the maximum system size one can treat.

In the present paper we investigate the replica-symmetric (RS) theory
of an infinite range spin-glass model for which the couplings
strengths are drawn from a Levy distribution \cite{CiBo}. The main 
characteristic of these distributions are power-law tails resulting in 
diverging moments. Spin glasses with Levy couplings are interesting
for several reasons. In real spin glasses with a random distribution 
of magnetic atoms in a non-magnetic host lattice the RKKY-interaction
give rise to a broad spectrum of interaction strengths, in particular for
low concentration of magnetic impurities. The coexistence of couplings
with vastly different strength is badly represented by a Gaussian
distribution as used in the SK model. Also, it is interesting to see
whether the concept of frustration which is central to the
understanding of spin glasses has to be modified for broad
distributions of coupling strengths. Moreover, in a completely connected
spin glass with Levy distributed couplings each spin will establish
${\cal O}(1)$ strong bonds with other spins whereas the majority of
couplings are weak, i.e. tend to zero for $N\to \infty$. Levy spin
glasses are hence intermediate between the classes represented by the VB
and SK model respectively and it is interesting to see how the spin
glass transition is influenced by the percolation of the strong bonds
on the one hand and the collective blocking of the many weak bonds on
the other hand. 

Levy spin glasses also pose new challenges to the theoretical analysis
because the diverging second moment of the coupling distribution
invalidates the central limit 
theorem which is at the bottom of many mean-field techniques. Related
issues of interest include quantum spin glasses with broad coupling
distribution \cite{KLRKI}, the spectral theory of random matrices with
Levy-distributed entries \cite{ranmat1,ranmat2}, and relaxation and
transport on scale-free networks \cite{AlBa}. It is also possible that
the peculiar properties of Levy distributions may facilitate 
mathematically rigorous investigations of spin glasses. In this
respect it is interesting to note that the properties of the 
Cauchy-distribution have recently enabled progress in the
mathematically rigorous analysis of matrix games with random pay-off 
matrices \cite{Roberts}.  

The Levy spin glass was investigated previously by Cizeau and Bouchaud
using the cavity method \cite{CiBo}. Complementary, our main emphasis
will be on the application of the replica method to the Levy 
spin glass. As noted also by Cizeau and Bouchaud a straightforward
implementation of the classical version of the replica method for
infinite range models \cite{EdAn} is impractical due to diverging
order parameters. It is, however, possible to
use a variant of the replica method that was developed to deal with
non-Gaussian local field distributions characteristic for diluted spin
glasses and complex optimization problems \cite{Remi}. Until now this
approach was used only in situations where the local field
distribution is inadequately characterized by its second moment alone
and higher moments of the distribution are needed for a complete
description. Here we show that the method may also be adapted to
situations where the moments may not even exist. 

The paper is organized as follows. After the precise definition of
the model in the next section we recall in section \ref{cavity} the
main steps of the RS cavity treatment performed by Cizeau and
Bouchaud. Section \ref{replica} comprises our replica analysis
including the results for the spin glass transition temperature and
the influence of a ferromagnetic bias in the coupling distribution. In
section \ref{numerics} we describe our numerical simulations and
compare their results for the transition temperature with our
analytical findings. Finally, section \ref{discussion} gives a short 
discussion of the results and points out some open problems.  


\section{The model}
\label{model}
We consider a system of $N$ Ising spins $S_i=\pm 1,\, i=1,...,N$ with
Hamiltonian 
\begin{equation}
  \label{eq:H}
  H(\{S_i\})=-\frac{1} {2 N^{1/\al}} \sum_{(i,j)} J_{ij} S_i S_j\; ,
\end{equation}
where the sum is over all pairs of spins. The couplings $J_{ij}=J_{ji}$
are independent, identically distributed random variables drawn from a
symmetric Levy distribution $P_\al(J)$. It is defined by its
characteristic 
function \cite{GnKo} 
\begin{equation}
  \label{eq:defP}
  \tilde{P}_\al(q):=\int dJ \; e^{-iqJ}\; P_\al(J)=e^{-|q|^\al} \; 
\end{equation}
with the real parameter $\al$, $0<\al<2$. A Gaussian distribution of
couplings as in the standard SK model is obtained in the limit $\al\to
2$. 

Levy distributions are
stable distributions which roughly means the following. If 
$x_i, i=1,...,N$ are independent random variables drawn from a Levy
distribution $P_\al(x)$ their sum, $z=\sum_i x_i$, is distributed
according to $P_\al(z/N^{1/\al})$, i.e. $z$ is also Levy distributed
with the same parameter $\al$, albeit with a width increased by a factor 
$N^{1/\al}$. Correspondingly the exchange fields
\begin{equation}
  \label{eq:defhex}
  h_i^{\mathrm{exch}} := \sum_j \frac{J_{ij}}{N^{1/\al}} S_j
\end{equation}
in a Levy spin glass are Levy distributed and the scaling of the
couplings with $1/N^{1/\al}$ ensures that they are of order~1 for
$N\to\infty$ such that the Hamiltonian (\ref{eq:H}) is extensive. 

>From the definition (\ref{eq:defP}) we also find the asymptotic
form of $P(J)$ for large $|J|$ to be  
\begin{equation}
  \label{eq:asyP}
  P(J)\sim \frac{1}{|J|^{\al+1}}\; .
\end{equation}
From this asymptotic behaviour and the interval of admissible values
of $\al$ it is clear that the second and higher moments of Levy
distributions do not exist. The long tail of the distribution also
implies that the largest among $N$ independent Levy variables is of
order $N^{1/\al}$, i.e. of  exactly the same order as their sum. The
sum is hence dominated by its largest summands. Each spin in a
Levy spin glass is therefore coupled to the majority of other 
spins by weak couplings of order $1/N^{1/\al}$ and to a few (${\cal
  O}(1)$ for $N\to\infty$) by strong bonds of order~1.  

The thermodynamic properties of the system are described by the
ensemble averaged free energy 
\begin{equation}
  \label{eq:deff}
  f(\beta):=-\lim_{N\to\infty}\frac{1}{\beta N}\overline{\ln
    Z(\beta)}\; , 
\end{equation}
with the partition function 
\begin{equation}
  \label{eq:defZ}
  Z(\beta):=\sum_{\{S_i\}}\exp(-\beta H(\{S_i\}))\; .
\end{equation}
Here $\beta$ denotes the inverse temperature and the overbar stands for the
average over the random couplings $J_{ij}$. 


\section{Cavity analysis}
\label{cavity}
The first statistical mechanics analysis of the Levy spin glass was
performed 15 years ago by Cizeau and Bouchaud \cite{CiBo} using a
variant of the cavity method. In the traditional form of the cavity
method for fully connected systems \cite{MPV} one considers a system
of $N$ spins $\{S_i\}$ in a pure equilibrium state and adds $N$ new
couplings $J_{0i}, i=1,...,N$ between these existing spins and a {\it
  cavity} which will later accommodate the $(N+1)\,$-st spin $S_0$. For
both the couplings $J_{ij}$ in the $N$-spin system and for the new
couplings one particular realization is considered. The exchange field
(\ref{eq:defhex}) in the cavity 
\begin{equation}
  h_0^{\mathrm{exch}}= \sum_{i=1}^N \frac{J_{0i}}{N^{1/\al}} S_i
\end{equation}
is then a random variable due to the thermal fluctuations of the
$S_i$. The clustering property of pure states of an equilibrium system
ensures that the 
connected correlation functions of the spins tend to zero for
$N\to\infty$ \cite{MPV} and therefore $h_0^{\mathrm{exch}}$ is a sum
over many, asymptotically independent random variables. If all $J_{0i}$
are of the same order of magnitude this implies a Gaussian
distribution of the cavity field uniquely characterized by its
variance. Further manipulations generate a self-consistent equation
for this variance from which all replica symmetric properties of
the system may be derived.  

In the case of a Levy spin glass, however, a typical realization of
the couplings $J_{0i}$ contains a few very large bonds. This
invalidates the central limit theorem (the Lindeberg criterion is not
fulfilled, see \cite{Feller}) and the cavity field distribution is not
Gaussian. The traditional form of the cavity method for infinite range
models is hence not applicable to the Levy spin glass. 

As observed by Cizeau and Bouchaud it is however possible to employ a
variant of the cavity method as later used also in the analysis of
spin systems on locally tree-like graphs \cite{MePa01} which is known to
physicist as Bethe-Peierls approximation \cite{Bethe,Peierls} and to
computer scientists as {\it belief propagation} \cite{YeFrWe}. This
method builds on the fact that with $S_i$ being a binary quantity its
marginal probability distribution 
\begin{equation}
  P(S_i)=\frac{1}{Z}\sum_{\{S_j\}_{j\neq i}}\; 
      \exp{(-\beta H(\{S_j\}))}
\end{equation}
can be parametrized by a single variable which we take to be the 
{\it local} field $h_i$ defined by  
\begin{equation}
  \label{eq:defh}
  h_i:=\frac{1}{\beta}\atanh \langle S_i \rangle\; .
\end{equation}
Accordingly we find 
\begin{equation}
  \label{eq:P_i}
  P(S_i)=\frac{e^{\beta h_i S_i}}{2\cosh(\beta h_i)}
\end{equation}
as well as 
\begin{equation}
  m_i:=\langle S_i \rangle =\tanh(\beta h_i)\; .
\end{equation}
The local field must not be confused with the exchange field
(\ref{eq:defhex}). Unlike the latter it is not thermally
fluctuating. If the cavity distribution is Gaussian the
thermal average of the exchange field coincides with the local field
\cite{MPV}. However, in the general case and in particular for the
Levy spin glass this does not hold. 

For the marginal distribution of the new spin $S_0$ we have
\begin{equation}
  P(S_0)=\frac{e^{\beta h_0 S_0}}{2\cosh(\beta h_0)}
\end{equation}
as well as
\begin{equation}
  P(S_0)=\sum_{\{S_i\}}\; \exp{(\beta S_0 \sum_{i=1}^N J_{0i} S_i)}\;
      P(\{S_i\})\; .
\end{equation}
Using the clustering property in the form 
\begin{equation}
  P(\{S_i\})=\prod_{i=1}^N \frac{e^{\beta h_i S_i}}{2\cosh(\beta h_i)}
\end{equation}
a straightforward calculation yields 
\begin{equation}
  \label{eq:resh_0}
  h_0=\frac{1}{\beta}\sum_{i=1}^N \atanh(\tanh(\beta h_i)
           \tanh(\beta\frac{J_{0i}}{N^{1/\al}}))\; .
\end{equation}
As observed by Cizeau and Bouchaud one may be tempted to expand in the
argument of the second $\tanh$ for $N\to\infty$ to find the familiar
expression 
\begin{equation}
  h_0=\sum_{i=1}^N \frac{J_{0i}}{N^{1/\al}} m_i\, .
\end{equation}
However, this would be unjustified since some of the
$J_{0i}/N^{1/\al}$ are not small. 

The local field $h_0$ as given by (\ref{eq:resh_0}) is a random
quantity both due to its dependence on the old couplings $J_{ij}$
determining the $h_i$ and on the new couplings $J_{0i}$. As long as
$|m_i|=|\tanh(\beta h_i)|<1$ the non-linearity in
(\ref{eq:resh_0}) suppresses the influence of the few large $J_{0i}$. 
As a consequence the second moment of the local field distribution $P(h_0)$
exists:
\begin{align}\nonumber
  Q:&=\overline{h_0^2}=\frac{1}{\beta^2} \overline{\sum_{i,j} 
      \atanh(\tanh(\beta h_i)\tanh(\beta\frac{J_{0i}}{N^{1/\al}}))\, 
      \atanh(\tanh(\beta h_j)\tanh(\beta\frac{J_{0j}}{N^{1/\al}}))}\\
      \nonumber
    &=\frac{1}{\beta^2} \sum_{i}
      \overline{\atanh^2(\tanh(\beta
        h_i)\tanh(\beta\frac{J_{0i}}{N^{1/\al}}))}\\\label{h1}
    &=\frac{N}{\beta^2} \int d h P(h)\int dJ P_\al(J) \;
      \atanh^2(\tanh(\beta h)\tanh(\beta\frac{J}{N^{1/\al}}))\,
      .
\end{align}
Cizeau and Bouchaud therefore argue that the central limit theorem may
be applied and that $P(h_0)$ is Gaussian \cite{CiBo,Cizeau}. Using
translational invariance of the ensemble averaged system (\ref{h1})
may then be written as a self-consistent condition for $Q$. Using 
(\ref{eq:defP}) this equation acquires the form 
\begin{equation}\label{eq:saddleQ}
  Q=\frac{C(\al)}{\beta^2} \int \frac{d h}{\sqrt{2\pi Q}} 
     \exp{(-\frac{h^2}{2 Q})} 
     \int \frac{d J}{|J|^{\al+1}} \;
      \atanh^2(\tanh(\beta h)\tanh(\beta J))\; ,
\end{equation}
where 
\begin{equation}
  \label{eq:defC}
  C(\al):=\frac{\Gamma(\al+1)\sin(\frac{\al}{2}\pi)}{\pi}>0
\end{equation}
is a numerical constant. 

Solving (\ref{eq:saddleQ}) numerically one
obtains $Q(\beta)$ from which the free energy and all thermodynamics
properties may be derived. In particular one easily verifies that the
paramagnetic state with $Q=0$ is always a solution. It is stable for
small $\beta$ and looses its stability at $\beta_c$ given by 
\begin{equation}
  \label{eq:T_c1}
  1=C(\al)\int \frac{d J}{|J|^{\al+1}} \, \tanh^2(\beta_c J)\; .
\end{equation}


\section{Replica theory}
\label{replica}

\subsection{General setup}
Within the replica approach we employ the replica trick \cite{EdAn} to
calculate the average in (\ref{eq:deff}), 
\begin{equation}
  \label{eq:replicatrick}
  \overline{\ln Z}=\lim_{n\to 0}\frac{\overline{Z^n}-1}{n}\; .
\end{equation}
As usual we aim at calculating $\overline{Z^n}$ for integer $n$
by replicating the system $n$ times, $\{S_i\}\mapsto \{S^a_i\},\,
a=1,...,n$, and then try to continue the results to real $n$ in order
to eventually perform the limit $n\to 0$.  

According to (\ref{eq:H}) and (\ref{eq:defZ}) the partition function
is a sum of exponential terms with the exponents linear in the
couplings $J_{ij}$. Due to the algebraic decay $P_\al(J)\sim
1/|J|^{\al+1}$ of the distribution $P_\al(J)$ for large $|J|$ the
average $\overline{Z^n(\beta)}$ hence diverges for real $\beta$ and we
cannot proceed in the usual way. 

On the other hand, for a purely
imaginary temperature, $\beta=-ik,\,k\in\mathds{R}, k>0$, we find from
the very definition of $P_\al(J)$, cf. (\ref{eq:defP}) 
\begin{equation}
  \label{eq:Zn}
  \overline{Z^n(-ik)}=\sum_{\{S_i^a\}} \exp
\Big(-\frac{k^\al}{2N}\sum_{i,j} \Big |\sum_a S_i^a S_j^a\Big |^\al 
     + {\cal{O}}(1)\Big) \; .
\end{equation}
Note that the scaling of the interaction strengths with $N$ used in
(\ref{eq:H}) makes the replica Hamiltonian extensive as it should be. 

As characteristic for a mean-field system the determination of
$\overline{Z^n}$ can now be reduced to an effective single site
problem. To this end we use the notation $\vs=\{S^a\}$ for a spin
vector with $n$ components. It is then convenient to introduce the
variables  
\begin{equation}
  \label{eq:c}
  c(\vs)=\frac1 N \sum_i\delta(\vs_i,\vs)\; ,
\end{equation}
describing the fraction of lattice sites that share one out of the $2^n$
realizations of the spin vector $\vs$ \cite{Remi}. Clearly 
\begin{equation}
  \label{eq:constraint}
  \sum_{\vs} c(\vs)=1\; .
\end{equation}
Because of the identity  
\begin{equation}\label{eq:sumf}
  \frac1 N \sum_{i,j} f(\vs_i,\vs_j)=
      N \sum_{\vs,\vs'} c(\vs)c(\vs') f(\vs,\vs')
\end{equation}
the exponent in (\ref{eq:Zn}) is seen to depend on the spin
configuration $\{\vs_i\}$ solely through the variables $c(\vs)$. In
order to transform the trace over $\{\vs_i\}$ into an integral over the 
$c(\vs)$ we only need to determine the number of spin configurations
that realize a given combination of $c(\vs)$. A standard
calculation yields 
to leading order in $N$
\begin{equation}
  \sum_{\{S_i^a\}} {\prod_{\vs}}'
    \delta\Big(\frac1 N \sum_i \delta(\vs_i,\vs)-c(\vs)\Big)
   =\exp\Big(-N\sum_{\vs}c(\vs)\ln c(\vs)\Big)\; 
\end{equation}
where the prime at the product denotes that the constraint 
(\ref{eq:constraint}) has to be taken into account. 

We may therefore write (\ref{eq:Zn}) in the form
\begin{equation}
  \overline{Z^n(-ik)}=
  \int\prod_{\vs}dc(\vs)\; \delta(\sum_{\vs}c(\vs)-1) \;
  \exp\Big(-N\Big[\sum_{\vs}c(\vs)\ln c(\vs)+\frac{k^\al}{2}
  \sum_{\vs,\vs'}c(\vs)c(\vs') |\vs\cdot\vs'|^\al\Big]\Big)\; .\label{eq:Zn2}
\end{equation}
In the thermodynamic limit, $N\to\infty$, the integral in
(\ref{eq:Zn2}) can be calculated by the saddle-point method. The
corresponding self-consistent equation determining the saddle-point
values $c^{(0)}(\vsi)$ of the $c(\vsi)$ is given by 
\begin{equation}
  \label{eq:saddle}
  c^{(0)}(\vsi)=\Lambda(n)\exp\Big(-k^\al
     \sum_{\vs}c^{(0)}(\vs)|\vs\cdot\vsi|^\al\Big)  \; ,
\end{equation}
where the Lagrange parameter $\Lambda(n)$ enforces the constraint 
(\ref{eq:constraint}). 

\subsection{Replica symmetry}
Within the replica symmetric approximation one assumes that the
solution of (\ref{eq:saddle}) is symmetric under permutations of the
replica indices. This implies that the saddle-point values $c^{(0)}(\vs)$
may only depend on the sum, $s:=\sum_a S^a$, of the components of the
vector $\vs$. After the limit $n\to 0$ is performed the function 
$c^{(0)}(s)$ can be related to the replica symmetric distribution
$P(h)$ of local magnetic fields (\ref{eq:defh}) via \cite{Remi} 
\begin{equation}
  \label{eq:Ph}
  c^{(0)}(s)=\int dh\; P(h)\; e^{-ikhs} \qquad\qquad 
  P(h)=\int \frac{ds}{2\pi}\; e^{ish}\; c^{(0)}(\frac s k) \; .
\end{equation}
In this way the self-consistent equation (\ref{eq:saddle}) may be
transformed to a self-consistent equation for $P(h)$. 

To proceed along these lines in the present case we use (\ref{eq:Ph})
in (\ref{eq:saddle}) and perform the following manipulations
\begin{align}\nonumber
k^\al \sum_{\vs} e^{-ikhs}|\vs\cdot\vsi|^\al&=
    \int dr\, |k r|^\al \, \sum_{\vs} \delta(r-\vs\cdot\vsi)\;
       e^{-ikhs}\\\nonumber
&=\int\frac{dr\,d\rh}{2\pi} |k r|^\al e^{ir\rh} \;
   \sum_{\vs}\exp\Big(-ikhs-i\rh\vs\cdot\vsi\Big)\\\nonumber
&=\int\frac{dr\,d\rh}{2\pi} |r|^\al   e^{ir\rh} \;
   \sum_{\vs}\prod_a \exp\Big(-iS^a(kh+k\rh\sigma^a)\Big)\\
&=\int\frac{dr\,d\rh}{2\pi} |r|^\al e^{ir\rh} \;
    [2\cos k(h+\rh)]^{\frac{n+\sigma}{2}}\;
    [2\cos k(h-\rh)]^{\frac{n-\sigma}{2}}\\\nonumber 
&\rightarrow \int\frac{dr\,d\rh}{2\pi} |r|^\al e^{ir\rh} 
\left[\frac{\cos k(h+\rh)}{\cos k(h-\rh)}\right]^{\frac\sigma 2}\; ,
\end{align}
where the limit $n\to 0$ was performed in the last line and 
$\sigma:=\sum_a\sigma^a$. Using $\Lambda(n)\to 1$ for $n\to 0$
\cite{Remi} we therefore find from (\ref{eq:saddle}) in the replica
symmetric approximation 
\begin{equation}\label{eq:h1}
  c^{(0)}(\sigma)= \exp\left(-\int dh P(h)\int\frac{dr\,d\rh}{2\pi}
    |r|^\al \exp\Big(ir\rh +\frac\sigma 2
    \ln\frac{\cos k(h+\rh)}{\cos k(h-\rh)}\Big)\right)\; .
\end{equation}
Using this result in (\ref{eq:Ph}) we get 
\begin{equation}
  \label{eq:Phrs}
  P(h)=\int\frac{ds}{2\pi}\exp
       \left(ish-\int dh' P(h')\int\frac{dr\,d\rh}{2\pi} |r|^\al
       \exp\Big(ir\rh+\frac s{2k}
       \ln\frac{\cos k(h'+\rh)}{\cos k(h'-\rh)}\Big)\right)\; .
\end{equation}
We are now in the position to continue this result back to real values
of the temperature by simply setting $k=i\beta$:
\begin{equation}
  \label{eq:resrs1}
  P(h)=\int\frac{ds}{2\pi}\exp
    \left(ish-\int dh' P(h')\int\frac{dr\,d\rh}{2\pi} |r|^\al
    \exp\Big(ir\rh-i\frac s{2\beta}
    \ln\frac{\cosh\beta(h'+\rh)}{\cosh\beta(h'-\rh)}\Big)\right)\; . 
\end{equation}
Finally the $r$-integral may be performed by using
\begin{equation}
  \int\frac{dr\,d\rh}{2\pi} |r|^\al e^{ir\rh} f(\rh)=
   -C(\al)\int\frac{d\rh}{|\rh|^{\al+1}}
   \begin{cases} 
        [f(\rh)-f(0)] & \text{if} \qquad 0<\al<1\\
        [f(\rh)-f(0)-\rh f'(0)] & \text{if} \qquad 1<\al<2 
        \end{cases}\; ,
\end{equation}
where $f'$ denotes the derivative of $f$ and $C(\al)$ is
defined in (\ref{eq:defC}). In our case we have $f(0)=1$ and $f'(0)=0$
as implied by $P(h)=P(-h)$. Hence no distinction between $\al<1$ and
$\al>1$ needs to be made. 

We therefore get finally the following self-consistent equation
for the replica symmetric field distribution $P(h)$ of a Levy
spin glass at inverse temperature $\beta$:
\begin{equation}
  \label{eq:resrs}
  P(h)=\int\frac{ds}{2\pi}\exp
    \left(ish+ C(\al)\int dh' P(h')\int\frac{d\rh}{|\rh|^{\al+1}} 
      \left[\exp\Big(-i\frac s{\beta}
      \atanh(\tanh\beta h'\tanh\beta\rh)\Big)-1\right]\right)\; .
\end{equation}
The structure of this equation is rather similar to the corresponding
equation for the VB model \cite{KaSo}. It is also interesting to look
at the second moment of $P(h)$ for which we find 
\begin{equation}
  \label{eq:varianceh}
  \langle h^2\rangle =\int dh P(h) h^2 =
   \frac{C(\al)}{\beta^2} \int d h P(h)
     \int \frac{d \rh}{|\rh|^{\al+1}} \;
      \atanh^2(\tanh(\beta h)\tanh(\beta \rh))\; 
\end{equation}
which is rather similar to (\ref{eq:saddleQ}). However, the $P(h)$
solving (\ref{eq:resrs}) is not Gaussian. This can be seen by
inserting a Gaussian $P(h')$ in the r.h.s. of (\ref{eq:resrs}) which
then gets not reproduced on the l.h.s. 

\subsection{Spin-glass transition}
The paramagnetic field distribution,  $P(h)=\delta(h)$, is always a
solution of (\ref{eq:resrs}). To test its stability we plug into the
r.h.s. of (\ref{eq:resrs}) a distribution $P_0(h)$ with a small second
moment, $\epsilon_0:=\int dh P_0(h)\, h^2 \ll 1$, calculate the l.h.s. (to be
denoted by $P_1(h)$) by linearizing in $\epsilon_0$ and compare the new
second moment, $\epsilon_1:=\int dh P_1(h)\, h^2$, with
$\epsilon_0$. We find $\epsilon_1>\epsilon_0$, {\it i.e.} instability of the
paramagnetic state, if the temperature $T$ is smaller than the
critical temperature $T_c(\al)$ given by 
\begin{equation}
  \label{eq:T_c}
  T_c(\al)=\left[C(\al)\int\frac{d y}{|y|^{\al+1}} \,
         \tanh^2(y)\right]^{1/\al}\; .
\end{equation}
This result coincides with (\ref{eq:T_c1}) of the cavity approach. To
determine the threshold value of $\beta$ at which 
the distribution of local fields develops a non-zero second
moment it is hence not decisive whether $P(h)$ becomes Gaussian or not. 
In the limit $\al\to 2$ (\ref{eq:T_c}) correctly reproduces the value
$T_c^{SK}=\sqrt{2}$ of the SK-model \cite{SK}.

It is interesting to compare the temperature for the spin-glass
transition with the temperature at which bonds satisfying $J_{ij}>T
N^{1/\al}$ start to percolate. From (\ref{eq:defP}) we find for the
fraction $c$ of these {\it strong} bonds per site 
\begin{equation}
  c=\frac{2 C(\al)}{\al}\; T^{-\al}\; .
\end{equation}
Since the strong bonds are distributed independently from each other
a giant component connected by these bonds appears for $c \ge 1$
\cite{ErRe}. The percolation temperature is hence given by
\begin{equation}
  \label{eq:T_p}
  T_p=\left(\frac{2 C(\al)}{\al}\right)^{1/\al}\; .
\end{equation}

\begin{figure}
  \begin{center}
    \includegraphics[width=0.6\textwidth]{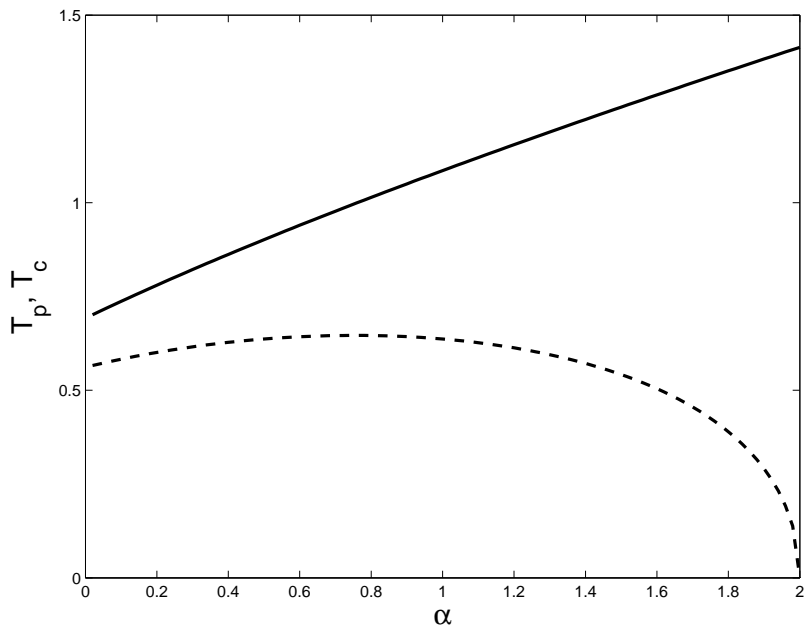}
  \end{center}
\caption{Spin glass transition temperature $T_c$ (full line) and
  percolation temperature $T_p$ (dashed line) of an infinite-range
  spin-glass with Levy-distributed couplings as function of the
  parameter $\al$ of the Levy distribution defined in
  (\ref{eq:defP}). For the scaling of the coupling strength with $N$
  as chosen in (\ref{eq:H}) there is a finite transition temperature
  for all values of $\al$.} \label{f.1}
\end{figure}

The dependence of $T_c$ and $T_p$ on $\al$ is displayed in
fig.~\ref{f.1}. The percolation temperature is always lower than the
spin-glass temperature as expected since a percolating backbone of
strong bonds is incompatible with a paramagnetic phase. On the other hand
the two temperatures never coincide which means that also the many weak
bonds contribute significantly to the spin-glass transition in Levy
spin glasses. The transition is therefore not a pure percolation
transitions. As can be seen 
from fig.~\ref{f.1} the difference between $T_c$ and $T_p$ decreases
with decreasing $\al$ in agreement with the fact that the tails of
$P(J)$ comprise a larger and larger part of the probability. 

\subsection{Asymmetric distribution of couplings}
\label{ferromagnetic bias}
The replica calculation described above may be generalized to the case
in which the distribution of couplings is not symmetric but shows a
ferromagnetic bias, $P_\al(J) \neq P_\al(-J)$.  
The couplings are then drawn from a Levy distribution centered at
$N^{\frac{1}{\alpha}-1} J_{0}$ where the $N$-dependence of the  
shift guarantees that the replica Hamiltonian remains extensive. 
Since a shift in the distribution amounts to a phase shift in the
characteristic function the exponent of the replicated
partition function is supplemented by 
\begin{eqnarray}
  -\frac{ik}{2 N}J_{0} \sum_{i,j} \vec S_i \cdot \vec S_j  =
     -N \frac{ik}{2 }J_{0} \sum_{ \vec S , \vec S'  } 
    c( \vec S) c( \vec S') \; \vec S \cdot \vec S' \; ,
\end{eqnarray}
where the identity (\ref{eq:sumf}) was used. The corresponding
saddle-point equation then reads 
\begin{eqnarray}
 c^{(0)}(\vsi)=\Lambda(n)\exp\Big(-k^\al
     \sum_{\vs}c^{(0)}(\vs)\; |\vs\cdot\vsi|^\al
     -ikJ_0 \sum_{\vec S }c^{(0)}(\vs)\;\vs\cdot\vsi  \Big) \; .
\end{eqnarray}
Using the RS ansatz we find for the new contribution 
\begin{align}\nonumber
-ikJ_0\int dh P(h) \sum_{\vec S }c^{(0)}(\vs)\; \vs\cdot\vsi &=
-kJ_0 \int dh P(h) \int\frac{dr\,d\rh}{2\pi} ir e^{ir\rh} 
\left[\frac{\cos k(h+\rh)}{\cos k(h-\rh)}\right]^{\frac\sigma 2}\\
\nonumber
&=-kJ_0\int dh P(h) \int d\rh \; \delta' (\rh) 
\left[\frac{\cos k(h+\rh)}{\cos k(h-\rh)}\right]^{\frac\sigma 2}\\
&=-kJ_0\sigma \int dh P(h) \tan(kh),
\end{align}
where the limit $n\to 0 $ was performed after the single-site trace
was completed. Performing the step back to real temperatures
$k=i\beta$ we get a self-consistent equation for the replica symmetric
field distribution. 
Since the field distribution is no longer symmetric for biased couplings,
it is necessary to distinguish between the cases $0<\alpha <1$ and
$1<\alpha <2$. In the former case we get 
\begin{eqnarray} \nonumber
   P(h)&=&\int\frac{ds}{2\pi}\exp
   \left(
   ish+ C(\al)\int dh' P(h')\int\frac{d\rh}{|\rh|^{\al+1}} 
   \left[\exp\Big(-i\frac s{\beta}
     \atanh(\tanh\beta h'\tanh\beta\rh)\Big)-1      \right]  \right.\\
  & & \qquad \qquad \qquad \quad \left. -isJ_0\int  dh' P(h')\tanh\beta h'
   \right) ,
\end{eqnarray}
while for the latter case the equation reads
\begin{eqnarray}\nonumber
   P(h)&=&\int\frac{ds}{2\pi}\exp
   \left(
   ish+ C(\al)\int dh' P(h')\int\frac{d\rh}{|\rh|^{\al+1}} 
   \left[\exp\Big(-i\frac s{\beta}
     \atanh(\tanh\beta h'\tanh\beta\rh)\Big)-1+i\rh s 
    \tanh( \beta h')  \right] \right. \\
   &&\qquad \qquad \qquad \quad \left.-isJ_0\int  
   dh' P(h')\tanh\beta h' \right).
\end{eqnarray}
From the structure of the self-consistent equation we again infer 
that the paramagnetic field distribution $P(h)= \delta(h)$ is always a
solution. To test its stability we use the same procedure as in the
previous section, taking into account that also a ferromagnetic
instability may occur. To this end we plug into the r.h.s.~of the
self-consistent equation a distribution $P_0(h)$ with mean
$\gamma_{0} := \int d h P_0(h) h  \ll 1$ and variance  
$\epsilon_{0}:= \int d h P_0(h) (h-\gamma_{0})^2 \ll 1$.
We calculate the l.h.s.~(to be denoted by $P_1(h)$) to the leading
order in the small parameters, and compare the resulting cumulants 
of the distribution $P_1$ with the corresponding quantities of the
distribution $P_0$. The phase transition from a paramagnetic to
a ferromagnetic state occurs, if 
\begin{eqnarray}
 \gamma_0 < \gamma_1 =  J_0\beta \gamma_0+
    \mathcal{O}(\gamma^2_0,\gamma_0 \epsilon_0, \epsilon_0^2) .
\end{eqnarray}
We therefore find an instability toward a ferromagnetic state at
$T_c^{\mathrm{FM}}= J_0$ which is independent of $\alpha$. 
The result for the spin glass transition temperature remains the same as 
in the unbiased case. 

\section{Numerical simulations}
\label{numerics}
In order to check our analytical results for the spin glass transition
temperature we have performed Monte Carlo simulations
\cite{newman1999,landau2000} using the parallel tempering approach
\cite{geyer1991,hukushima1996}. For a given realization $\{J_{ij}\}$
of the disorder, $K$ independent configurations $\{S_i^k\}$
($k=1,\dots,K$) are simulated at $K$ different temperatures
$T_1<T_2<\ldots<T_K$, i.e.\ $\{S_i^1\}$ at $T_1$, $\{S_i^2\}$ at $T_2$
etc \footnote{The upper 
  index $k$  of the spin variables must not be confused with the
  replica index of section \ref{replica}.}. One step of the
simulation, i.e.\ one Monte Carlo sweep, consists of the following
steps: 

\begin{itemize}
\item For each of the configurations $k=1,\dots,K$, 
one sweep of {\em local Metropolis
steps} is performed. Each sweep
consist of $N$ times selecting a spin $i_0\in\{1,\ldots,N\}$ randomly 
(uniformly). For each selected spin, the
energy difference $\Delta E$ between the current configuration $\{S_i^k\}$
and the configuration where just spin $S_{i_0}^k$ is flipped 
($S_{i_0}^k\to -S_{i_0}^k)$ is calculated:
$\Delta E = H(\{S_i^k\})-H(\{S_i^k|-S_{i_0}^k\})$.
The flip of spin $S_{i_0}^k$ is actually performed with the Metropolis
probability $p_{\rm flip}=\min\{1,\exp(-\Delta E/T_k)\}$, otherwise
the current configuration remains unaltered.

\item $K-1$ times an {\em exchange step} is tried: 
A temperature $k_0\in\{1,\ldots,K-1\}$ is selected randomly, each temperature
with the same probability $1/(K-1)$. The energy
difference $\Delta E_{\rm exch} = H(\{S_i^{k_0}\}) - H(\{S_i^{k_0+1}\})$
between the configurations at neighboring 
temperatures $T_{k_0}$ and $T_{k_0+1}$ is calculated. The two
configurations $\{S_i^{k_0}\}$ and $\{S_i^{k_0+1}\}$ are exchanged
with probability $p_{\rm exch}=\min\{1, \exp(-\Delta E_{\rm exch}
(1/T_{k_0}-1/T_{k_0+1}))\}$. In this way, the configurations perform a
random walk in temperature space and can visit all temperatures $T_k$.
\end{itemize}

Furthermore, for each temperature, we simulate two independent sets
of configurations  $\{S_i^k\}$, $\{\tilde {S}_i^k\}$ which allows for
a simple calculation of the overlap 
\begin{equation}
  \label{eq:defq}
  q=\frac{1}{N}\sum_i S_i \tilde{S}_i
\end{equation}
at each temperature $T_k$. From this overlap we calculate the Binder
cumulant 
\begin{equation}
  \label{eq:defB}
  B_N(T)=\overline{\frac1 2
   \left(3-\frac{\langle q^4\rangle}{\langle q^2\rangle^2}\right)}
\end{equation}
for all $T_k$ and various values of $N$. The critical temperature is
then determined from the intersection points of the lines $B_N(T)$ for
different values of $N$.  

\begin{figure}
\includegraphics[width=0.8\linewidth]{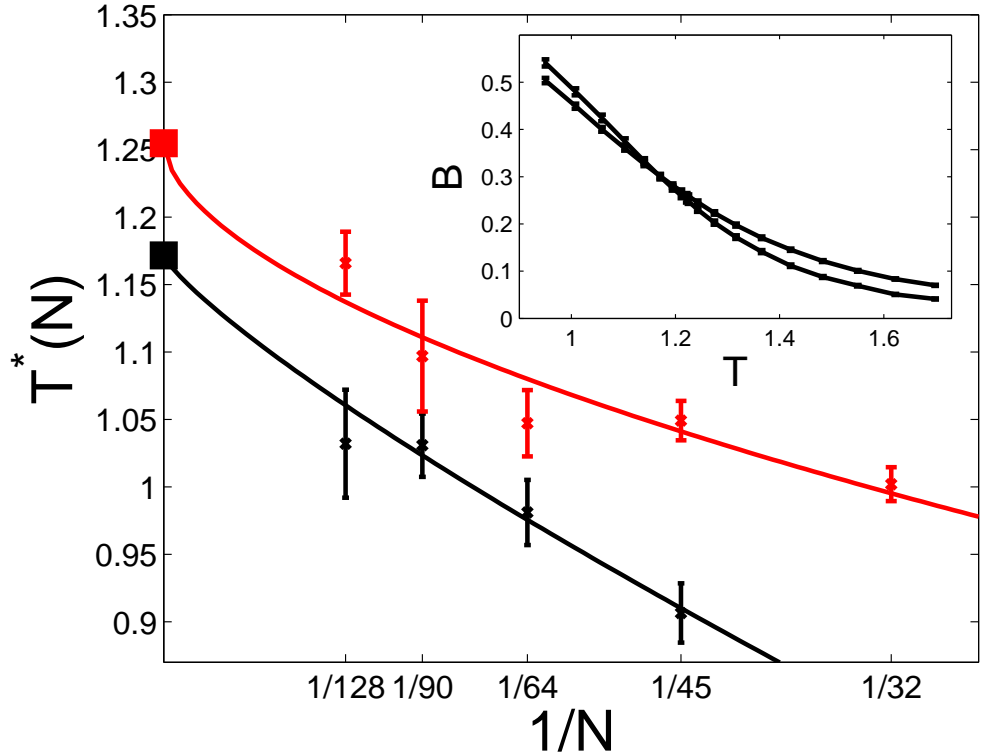}
\begin{center}
\centering\caption{
Transition temperatures $T^\star(N)$ for $\al = 1.25$ (black) and 
$\al = 1.5$ (red)  as a function of the inverse system
size. $T^\star(N)$ is determined from the intersection points of the
Binder parameters of the overlap $q$ for system sizes 
$N$ and $2N$ as shown in the inset for $N=128$ and $N=256$
($\al=1.5$). The lines show fits for the finite-size scaling of the
form $T^\star(N)=T_c+a N^{-b}$. Here $T_c(\al=1.25)\simeq 1.172$ and 
$T_c(\al=1.5)\simeq 1.254$ respectively indicated by
the boxes on the left of the figure are the analytical values for
the critical temperatures as given by (\ref{eq:T_c}) whereas $a$ and $b$
are fit parameters. As can be seen the numerical results and the
analytical values are compatible with each other.
}
\label{f2}
\end{center}
\end{figure}

We have found that the traditional single-spin local update works very
well for large values of $\alpha\ge 1.5$. For smaller values, the
probability 
that bonds with a very large magnitude (e.g.\ $|J_{ij}|>10$) appear in a 
realization becomes significant. A spin, which is adjacent to such a
bond, will satisfy such a bond on all timescales, for the range of
temperatures studied here. Hence, the spin will be frozen under
single-spin-flip dynamics. For this reason, we have extended the local
update by a {\em cluster flip}: In advance, all large bonds with
$|J_{ij}|\ge J_{\max}$ are determined. 
Next, we calculate the
maximal components of sites connected by these large bonds. The
cluster flip consist of an attempt to flip a randomly chosen 
cluster, i.e.\ all spins of the cluster simultaneously, with the usual
Metropolis $p_{\rm flip}$ probability as stated above, where
$\Delta E = H(\{S_i^k\})-H(\{S_i^k|{\rm cluster\; flipped}\})$.
Note that the clusters contain only spin indices, i.e.\ are indpependent
of the actual relative orientations of the spins, since these
 might change during
the simulation by other update steps, e.g.\ the standard 
single-spin-flip step.

In case of a single chosen value of $J_{\max}$, if $J_{\max}$ is large,
then the clusters will be small, which might  lead, in some cases,
not to frozen single spins but to some practically frozen clusters. 
On the other hand, if $J_{\max}$ is small,
the clusters will be large, hence the spins inside a cluster are frozen
relative to each other. To avoid these problems, 
we have generated, before the actual simulation starts,
 several sets $C_n$ of  clusters
for different values of $J^n_{\max}$. We started at $J^1_{\max}=2 T_K$ to 
obtain $C_1$.  Then we increment $J_{\max}$ iterativly by 1. A new
set $C_n$ is stored, if it differs from the previous set $C_{n-1}$.
This is continued until $C_{n_{\max}}$ consists only of clusters of size 2.
During the simulation of the single configurations, each time the
local Metropolis step is chosen with probability $p=0.8$ and a cluster
attempt with probability $1-p=0.2$. 
Note that for the Metropolis step still all
spins are considered for single-spin flips, independent on how the
clusters look like. Hence, large bonds which
are not satisfied will become satisfied in this way, and also there
is a small probability that bonds with a large magnitude  become
unsatisfied during the simulation. Hence, ergodicity is guaranteed.
For the cluster attempt, one set $C_n$
of clusters is selected randomly (all with the same probability), and
from the set one cluster, again with equal probability.
Hence, detailed balance is fulfilled. 

After checking our code by reproducing the known result
$T_c(\al=2)=\sqrt{2}$ for the SK model we have investigated the cases
$\al=1.25$ and $\al=1.5$ in more detail. Guided by the analytical result
(\ref{eq:defC}) for  $T_c(\al=1.25)\simeq 1.172$ and 
$T_c(\al=1.5)\simeq 1.254$ we chose in both cases 19
temperatures in the range $[0.87:2.0]$. 
The temperatures $T_i$ are
determined such that for the largest system size $N=256$ the average
acceptance rate of the exchange steps is at least 0.5 for
all pairs of neighboring temperatures. For all system sizes,
the same set of temperatures is used, which allows for a
better comparison of the results.

At the beginning of the simulation all configurations are random. We
equilibrate the system until the squared overlap $q^2$ as a function
of time, averaged over 
the last half of the simulation, has become independent of the number
of Monte Carlo sweeps for all temperatures. Furthermore, we verify
that the distribution of overlaps measured during this period is
symmetric with respect to $q=0$. For the case $\alpha=2.0$, we have
additionally employed the equilibration criterion from
\cite{katzgraber2001} and verified that the above listed criteria 
are compatible with it. 

After equilibration, spin configurations are stored for later analysis
at all temperatures every $\Delta t$ Monte Carlo sweeps. $\Delta t$ is
chosen such that it corresponds to the typical time one configuration
needs to walk in temperature space from the lowest temperature $T_1$
to the highest $T_K$ and back to $T_1$. Since at the highest temperature,
well above the phase transition temperature, the configurations forget
their history at low temperatures, 
the stored configurations are  statistically  independent. Typical values
for $\Delta t$ range from $\Delta t = 150$ ($N=32$) to $\Delta
t=250$ ($N=256$). For each realization, we sample 1000
configurations and average for each system size over 1000
realizations. The results for $T_c$ obtained in this way are
compatible with the theoretical result as shown in fig.~\ref{f2}.


\section{Discussion}
\label{discussion}
Infinite-range spin glasses with Levy-distributed couplings are
interesting examples of disordered systems. Due to the long
tails in the distribution of coupling strengths they 
interpolate between systems with many weak couplings per spin as the
Sherrington-Kirkpatrick model and systems with few strong couplings
per spin as the Viana-Bray model. 
The broad variations in coupling strengths brought about by the
power-law tails in the Levy distribution violate the Lindeberg
condition for the application of the central limit theorem and give
rise to non-Gaussian cavity field distributions with diverging
moments. In the present paper we have shown that it is nevertheless
possible to derive the replica symmetric properties of the system in a
compact way by using the replica method as developed for the treatment
of strongly diluted spin glasses and optimization problems
\cite{Remi}. This approach focuses from the start on the complete
distribution of fields rather than on its moments. 

The central result of our analysis is the self-consistent equation
for the distribution of local fields, $P(h)$, as given by
eq.~(\ref{eq:resrs}). From this equation the expression (\ref{eq:T_c})
for the critical temperature of the spin glass transition may be
derived. In Levy spin glasses there is for all temperatures a fraction
of strong bonds per site which cannot be broken thermally. Comparison
of the spin-glass transition temperature
with the temperature at which these strong bonds start to percolate
through the system reveals that the spin-glass transition in a Levy
glass is not a pure percolation transition. The contribution of the
many weak couplings cannot be neglected and becomes increasingly
important as the parameter $\al$ in the Levy distribution approaches
the limit $\al=2$ corresponding to the SK model. 

Our results show similarities and differences with those of the cavity
analysis of Cizeau and Bouchaud \cite{CiBo}. The results for the
critical temperature are the same because the expressions for the
second moment of the local field distribution coincide. However, we do
not find a Gaussian distribution of local fields for $T<T_c$ as
assumed by Cizeau and Bouchaud on the basis of the cavity expression
(\ref{eq:resh_0}). From the numerical solution of TAP equations for the
SK model it is known that for all $T<T_c$ a certain fraction of 
local magnetizations $m_i=\tanh\beta h_i$ are extremely near to 1
\cite{Timo}. As this seems likely to be the case in Levy spin glasses
as well it is conceivable that the distribution of $h_i$ in
(\ref{eq:resh_0}) is such that the Lindeberg criterion is again
violated and that the central limit theorem may not be applicable.  

Several open questions may be addressed in forthcoming work in order
to completely characterize the properties of Levy spin
glasses. First the self-consistent equation for $P(h)$ should be
solved, either numerically or analytically in limiting cases. Building
on these results the replica symmetric picture of the low-temperature
phase may be completed and compared with the findings from the cavity
approach. Since replica symmetry is certainly broken at low
temperature a stability analysis of the RS saddle point
(\ref{eq:saddle}) needs to be performed and it is to be checked
whether the deAlmeida-Thouless temperature $T_{\mathrm{AT}}$ is indeed
smaller than $T_c$ as found within the cavity approach. Finally the
structure of the solution with broken replica symmetry is to be
elucidated. Within the replica approach adopted in the present paper
this is known to be very complicated such that for this task a cavity
analysis looks more promising. Finally, improved numerical simulations
will contribute to a better understanding of the intricate properties
of Levy spin glasses.

\acknowledgments We would like to thank Timo Aspelmeier, Daniel
Grieser, R\'emi Monasson and Martin Weigt for clarifying
discussions. A.~E. is deeply indebted to Thomas Nattermann for 
introducing him to the statistical mechanics of disordered systems
25 years ago.

\end{document}